\documentstyle[prl,aps,twocolumn,graphicx]{revtex}
\input{amssym}

\title{ Learning multilayer perceptrons efficiently
      }

 \author{C. Bunzmann, M. Biehl , R. Urbanczik  \\
        Institut f\"ur theoretische Physik\\
        Universit\"at W\"urzburg \\
        Am Hubland\\
        D-97074 W\"urzburg \\
        Germany
       }

\newcommand{\R}{I\!\!R}
\newcommand{\sign}{ \mbox{\rm sgn} }
\newcommand{\erf}{ \mbox{\rm erf} }
\newcommand{\ext}{ \mbox{\rm extr} }
\newcommand{\exta}[1]
     {{\renewcommand{\arraystretch}{0.75} \begin{array}[t]{c} 
       \ext \\ {\scriptstyle #1}
     \end{array}}}

\newcommand{\Tr}{{ \mbox{\rm Tr}} }
\newcommand{\epsg} {\epsilon_{\rm g}}
\newcommand{\alcK} {\alpha_{\rm c}}

\newcommand{\half}{{\frac{1}{2}}}

\newcommand{\La}{\left\langle}
\newcommand{\Ra}{\right\rangle}
\newcommand{\cut}[1]{}

\begin{document}
\maketitle

\begin{abstract}
 A learning algorithm for multilayer perceptrons is presented which is
  based on finding the principal components of a correlation matrix
 computed from the example inputs and their target outputs. For large 
 networks our procedure needs far fewer examples to achieve good 
 generalization than traditional on-line algorithms. 
\end{abstract}

\ 


\


Multilayer neural networks achieve the ability to approximate
any reasonable function arbitrarily well \cite{Cyb89} by interconnecting
sufficiently many architecturally identical processing elements, the
perceptrons. This replication of identical elements invariably
leads to symmetries in the multilayer architecture which need 
to be broken during training to achieve good generalization \cite{Schw93a}.
In the dynamics of gradient based training procedures
the symmetries give rise to suboptimal fixed points and slow convergence.
In particular this holds for stochastic gradient descent methods which
to date have been the method of choice for training large networks 
because, at each time step, the update of the network parameters is 
based on the presentation of only 
a single training example and hence computationally cheap.
Such stochastic gradient descent methods have
been intensively studied in the framework of on-line learning, where
each of the training examples is used just once by the network \cite{Saa95}. 
This is attractive
because there is no need to store the entire set of training examples,
and the approach also works for nonstationary problems. However,
in the on-line framework the number of required training examples is coupled 
to the slow temporal evolution caused by the symmetries and thus
unreasonably large training sets are necessary. 
In fact the ratio of the number of examples needed for
good generalization to the number
of free parameters in the network diverges with the network size \cite{Bie96}.
While there have been investigations into optimizing the on-line
dynamics \cite{Saa97,Vic97,Rat98}, these have not lead to practical 
algorithms since the optimized procedures assume that the symmetry breaking
provided by the initial conditions is macroscopic and known to the learning
algorithm.

In this letter we present a learning algorithm 
which has many of the attractive features of the traditional
on-line procedures but yields good generalization using a much smaller 
number of training examples.
Further an exact analysis of the algorithm's performance shows
that the ratio of required examples to the number of free parameters
in the networks does stay finite in the thermodynamic limit.

The multilayer architecture we analyze is a
committee machine with $K$ hidden units defined by
\begin{equation}
\tau(\xi) = g\left(K^{-1/2} \sum_{i=1}^K h(B_i^T \xi)\right), \label{comm}
\end{equation}
where $\xi \in \R^N$ is the input and 
the $B_i\in \R^N$ are the unknown parameter vectors. 
The goal of learning is to estimate 
these parameter vectors, which we shall refer to as teacher vectors,
from a training set of
$P$ examples  $(\xi^\mu,\tau(\xi^\mu))$ of the input/output relationship.
We shall initially focus on regression problems and later
sketch the modifications to our procedure needed for classification tasks.
For regression, the output function $g$ is usually assumed invertible,
and this easily reduces to the linear case by applying the inverse function
to the target outputs. So in this case we simply assume $g(x)=x$.
For brevity we also assume that the $B_i$ are orthonormal.

In its simplest form our
procedure can be seen as generalizing  Hebbian learning.  There
the parameter vector $\bar{J}^P$ of a simple perceptron approximating 
the target function $\tau(\xi)$ is obtained as the average 
\begin{equation}
\bar{J}^P = P^{-1}\sum_{\mu=1}^P \tau(\xi^\mu) \xi^\mu. \label{hebb}
\end{equation}
Our basic observation is that when $\tau(\xi)$
is given by a multilayer perceptron it is important to not only consider
the empirical mean of the distribution of $\tau(\xi^\mu)\xi^\mu$ as in 
plain Hebbian
learning but also its correlation matrix.  In particular some of the 
eigenvectors of the correlation matrix correspond to the parameter vectors 
$B_i$ and thus the supervised learning task is mapped onto the
well understood problem of principal component analysis. While this mapping
by itself does not solve the original problem, it does provide a crucial
reduction in its dimensionality, such that the remaining problem
becomes almost trivial in the thermodynamic limit $N\rightarrow\infty$.

We shall consider the general correlation matrix
\begin{equation}
C^P = P^{-1}\sum_{\mu = 1}^P F\left(\tau(\xi^\mu)\right) \xi^\mu {\xi^\mu}^T
\end{equation}
where the simple choice $F(x)=x^2$  for the weight function $F$ yields the
analogy to Hebbian learning. Assuming the components
of the input vectors $\xi^\mu$ to be independent Gaussian random variables
with zero mean and unit variance, it is straightforward to analyze
the spectrum of $C^P$ in the limit $P\rightarrow\infty$. 
The spectrum then only has
\newcommand{\lamorth}{\lambda_0}
\newcommand{\lamav}{\bar{\lambda}}
\newcommand{\lamdiff}{\lambda_\Delta}
\newcommand{\tauyi}{\tau_y}
three eigenvalues $\lamorth,\lamav$ and $\lamdiff$.  The degeneracy 
of $\lamorth$ is $N-K$ and any vector orthogonal to all teacher vectors
$B_i$ is an eigenvector. The eigenvalue $\lamav$ has the single eigenvector 
$\bar{B} = K^{-1/2}\sum_{i=1}^KB_i$; this eigenvector is of little
interest since for large $P$ one also has $\bar{B} \propto \bar{J}^P$, 
and it is thus simpler to use Hebb's rule (\ref{hebb}) to estimate
$\bar{B}$.  The important
eigenspace is the one of $\lamdiff$ since it is spanned by the  
$K-1$ vectors  $B_1-B_j$ ($j=2,\ldots,K$).
The three eigenvalues can be written as averages over independent, 
zero mean, unit variance, Gaussian random variables $y_i$. For instance
one obtains 
$\lamdiff  = \frac{1}{2}\La F(\tauyi)(y_1-y_2)^2 \Ra_y$
\cut{
\begin{equation}
\lamorth  = \La F(\tauyi) \Ra_y, \quad
\lamav    = \frac{1}{K}\La F(\tauyi)(\sum_{i=1}^K y_i)^2 \Ra_y, \quad
\lamdiff  = \frac{1}{2}\La F(\tauyi)(y_1-y_2)^2 \Ra_y, \label{eigs}
\end{equation}
}
where $\tauyi = K^{-1/2}\sum_{i=1}^K h(y_i)$. The activation function
$h$ is sigmoidal (odd, monotonic and bounded),
and when stating specific numerical values we will always assume  
$h(y) = \erf(y)$. 
Then for the choice $F(x)=x^2$ one finds the ordering
$\lamav > \lamorth > \lamdiff$ and 
$\lamorth - \lamdiff =  \frac{1}{K}\frac{8}{3 \pi}(1 - \frac{1}{\sqrt 5}  )$.

For a finite number $P$ of training examples the degeneracy in the spectrum
is broken by random fluctuations. But a computation of
the orthonormal eigenvectors $\Delta_j^P$ corresponding to the $K-1$ smallest 
eigenvalues of $C^P$ nevertheless yields an estimate of the space spanned
by the  difference vectors $B_1-B_j$. To measure the success of approximating
this space, we introduce the overlap
\begin{equation}
\rho = (K-1)^{-1/2}\Tr( \Delta^T B B^T \Delta )^{1/2}\,,
\end{equation}
where B  is the matrix $(B_1,\ldots,B_K)$ of the teacher vectors and 
$\Delta = (\Delta_1^P,\ldots,\Delta_{K-1}^P)$. 
This is a sensible measure because
$\rho$ is invariant with respect to the choice of an orthonormal basis
of the space spanned by the $\Delta_j^P$ and since it attains its
maximal value of
$1$ iff all $\Delta_j^P$ lie in the space spanned by the 
$B_i$. Simulations showing the overlap $\rho$ as 
function of the number of examples per free parameter, $\alpha = P/(KN)$, are
depicted in Fig. 1. They indicate a second order phase transition from zero
to positive $\rho$ at a critical value $\alcK$.
The evolution of $\rho$ can be calculated exactly in the thermodynamic
limit $N\rightarrow\infty$.  But since up to now  we are only estimating
the space spanned by the $B_i$, instead of the vectors themselves, 
we address this problem first and defer the calculation of $\rho$.

Using the above eigenspace procedure, the $B_i$ can be 
approximated by linear combinations of the $\Delta_j^P$ and $\bar{J}^P$.
Thus the original  $KN$ dimensional problem is reduced to 
$K^2$ dimensions and this is much easier for large $N$.
To exploit the dimensionality reduction, we write the possible linear
combinations as $\tilde{\Delta}\Gamma$ where 
$\tilde{\Delta}=(\bar{J}^P,\Delta_1^P,\ldots,\Delta_{K-1}^P)$, $\Gamma$
is a $K$ by $K$ matrix of tunable parameters,  and we define a student network
$\sigma_\Gamma$ in the restricted space via 
$
\sigma_\Gamma(\xi) = g\left(K^{-1/2} \sum_{i=1}^K 
                        h\left((\tilde{\Delta}\Gamma)_i^T\xi\right)\right)
$.
We want to choose $\Gamma$ to minimize the generalization error
$
\epsg = \La\half (\tau(\xi)-\sigma_\Gamma(\xi))^2 \Ra_\xi 
$
and to this end $\hat{P}$ additional examples
$(\hat{\xi}^\nu,\tau(\hat{\xi}^\nu))$, $\nu=1,\ldots,\hat{P}$, are used.
To simplify the theoretical analysis, the additional examples should
be picked independently of the examples used to obtain $\tilde{\Delta}$.
We then apply the standard on-line gradient descent procedure 
$\Gamma^{\nu+1} =  \Gamma^{\nu} - 
\eta \nabla_\Gamma 
\left(\tau(\hat{\xi}^\nu)-\sigma_{\Gamma^\nu}(\hat{\xi}^\nu)\right)^2$.
However, by choosing a scaling of the learning rate $\eta$ such that
$\eta \ll 1$ for large $N$, the stochasticity drops out of the procedure 
\cite{Ama67} and in the restricted space it performs
gradient descent in $\epsg$. Further we can scale
$\hat{P}$ such that $\eta \hat{P} \gg 1$ and then $\Gamma^{\hat{P}}$ will
be a minimum of $\epsg$ for large $N$. Finally both scaling conditions
can be satisfied while observing $\hat{P} \ll N$, so that the required 
number of 
additional examples is negligible compared to the size of the first
training set. Note that thanks to the reduced dimensionality
the details of the scaling of $\eta$ and $\hat{P}$
only affect finite size effects and not the performance in the
thermodynamic limit. Simulations combining the two stages of our algorithm
(Fig. 1) show a steady decay in $\epsg$.

We now turn to the theoretical calculation of $\rho$ in the important
first stage of the algorithm. 
The smallest eigenvalue of $C^P$ can be found by 
minimizing $J^TC^PJ$ under the constraint $|J|=1$. Hence we consider the
partition function
$
Z = \int dJ \exp(-\beta P  J^TC^PJ)
$
where the integration is over the $N$-dimensional unit sphere. 
For large $N$ the typical 
properties of the minimization problem are found by calculating
the training set average $\La \ln Z \Ra$ and taking the limit 
$\beta\rightarrow\infty$. Using replicas, one introduces the
$K$ dimensional order parameter $R$ , the typical overlap $B^T J$
of the teacher vectors with  a vector $J$ drawn from the Gibbs distribution 
$Z^{-1}\exp(-\beta P  J^TC^PJ)$, and further the replica symmetric
scalar order parameter
$q$ which is the overlap ${J^1}^TJ^2$ of two vectors drawn from this 
distribution. 
\cut{
...................................
Then $N^{-1} \La \ln Z \Ra$ is obtained as the extremum w.r.t
$R$ and $q$ of the function
\begin{eqnarray}
S = && 
  -\frac{\alpha K}{2} 
        \La \ln\frac{F(\tauyi)}{G_\chi(\tauyi)}+
             2\beta G_\chi(\tauyi)(R^T(y-R) + q)
                   \Ra_y \nonumber \\
       &&+        \half\left( \ln(1-q) + \frac{q-R^TR}{1-q} \right) \, ,
\end{eqnarray}
where $G_\chi(\tauyi) =  F(\tauyi)/(1 + 2 \chi F(\tauyi))$ and 
$\chi=\beta(1-q)$.
....................................
}
For $P > N$ the correlation matrix
$C^P$ is nonsingular, $q$ approaches $1$ with increasing
$\beta$, and the relevant scaling is $\chi =  \beta(1-q) = {\cal O}(1)$.
Then for the quenched 
average of $\ln Z$ one finds in the limit $\beta\rightarrow\infty$ 
\begin{equation}
\frac{\La \ln Z \Ra}{\beta N} = \exta{R,\chi} R^T A(\alpha,\chi) R + 
                                 a(\alpha,\chi)\,. \label{ext1}
\end{equation}
Here 
$a(\alpha,\chi) = -\alpha K \La G_\chi(\tauyi) \Ra_y + \frac{1}{2 \chi}$,
the $K$ by $K$ matrix $ A(\alpha,\chi)$ has entries 
\begin{equation}
A_{jk}(\alpha,\chi) =  -\alpha K \La  
  G_\chi(\tauyi)(y_jy_k-\delta_{jk}) \Ra_y 
     - \frac{\delta_{jk}}{2 \chi}\,,
\end{equation}
and $G_\chi(\tauyi) =  F(\tauyi)/(1 + 2 \chi F(\tauyi))$.
Since Eq. (\ref{ext1}) is quadratic in $R$ the extremal problem can only
have a solution $R \neq 0$ if the matrix $A$ is singular. From the symmetries
one easily obtains that $A$ has just two eigenvalues.
The first can be written as $A_{11}-A_{12}$, it
is the relevant eigenvalue in our case
\cite{foot1}
and its eigenspace is spanned by the vectors $e_1-e_j$ ($j = 2,\ldots,K$),
where $e_1,\ldots,e_K$ denote the standard basis of $\R^K$. This
degeneracy shows that the difference between the $K-1$ smallest
eigenvalues of the correlation matrix vanishes for large $N$. 
So the simple procedure
of analyzing the properties of the single vector $J$ minimizing
$J^T C^P J$,  in fact yields the properties of the $K-1$ eigenvectors
vectors of $C_P$ with smallest eigenvalues in the thermodynamic limit.
 
Due to the degeneracy,  we can reparametrize (\ref{ext1}) 
setting $R = \rho (e_1-e_j)/\sqrt{2}$ and obtain an extremal problem
with only two variables $\rho$ and $\chi$.
\cut{
can reduce the number of parameters in the 
extremal problem (\ref{ext1}) by   parametrizing $R$ as 
$R = \rho (e_1-e_j)/\sqrt{2}$ to obtain:
\begin{equation}
\frac{\La \ln Z \Ra}{\beta N} = 
\exta{\rho,\chi} 
 -\frac{\rho^2}{2} 
  \left(\alpha K\La G_\chi(\tauyi)((y_1-y_2)^2-2) \Ra_y - 
  \frac{1}{\chi}\right)  + a(\alpha,\chi) \label{ext2}
\end{equation}

}
Note that $\rho$ is indeed the parameter introduced in the analysis of the 
numerical simulations. 
Its evolution is now obtained by solving (\ref{ext1})
and this 
confirms the continuous phase transition found in the simulations
from $\rho=0$ to positive $\rho$ at a critical value $\alcK$. 
For $K=2$ one finds $\alcK = 4.49$
and for $K=3$ the result is $\alcK = 8.70$. As shown in Fig. 1
beyond the phase transition there is excellent agreement between the 
theoretical prediction and the simulations. 

To obtain generic results for
large $K$ note that the contributions of
$y_1$ and $y_2$ to the target output $\tauyi$ will be small in this limit.
Decomposing $\tauyi$ as 
$\tauyi = \tauyi^* + \delta_y/\sqrt{K}$ where $\delta_y = h(y_1)+h(y_2)$, and
expanding $G_\chi(\tauyi)$ up to second order for large $K$ simplifies 
Eq. \ref{ext1} to: 
\begin{eqnarray}
\frac{\La \ln Z \Ra}{\beta N}  = 
\exta{\rho,\chi}&& 
 -\frac{\alpha\rho^2}{4} 
  \La G''_\chi(\tauyi^*)\Ra_y  \La\delta_y^2 ((y_1-y_2)^2-2) \Ra_y \nonumber \\
 &&-\alpha K \La G_\chi(\tauyi) \Ra_y
   + \frac{1-\rho^2}{2 \chi}\;. \label{ext4}
\end{eqnarray}
On the one hand, applying the central limit theorem,
the multiple integrals $\La G''_\chi(\tauyi^*)\Ra_y$ and
$\La G_\chi(\tauyi) \Ra_y$ can now be replaced by a single average,
on the other hand  the structure of (\ref{ext4}) shows that 
$\chi$ approaches zero with increasing $K$. These observations yield the very
simple result that $\rho^2 = 1 - \alcK/\alpha$ for large $K$ and
$\alpha > \alcK$.  The value of $\alcK$ is obtained as
\begin{equation}
\alcK = \frac{4K \La F^2 ( \mu z ) \Ra_z}
             { \La F''( \mu z )\Ra_z^2
                \left( \La z^2 h^2(z) \Ra_z - \mu^2 - 
                 2 \La z h(z) \Ra_z^2 \right)^2}  \label{alck}
\end{equation}
where $\mu^2 = \La h^2(z)\Ra_z$, and the distribution of $z$ is Gaussian
with zero mean and unit variance. 
It is now straightforward to derive the optimal choice
of $F$ from Eq. (\ref{alck}) by noting that in the denominator  
$\La F''(\mu z)\Ra_z = \mu^{-2} \La (z^2-1)F(\mu z)\Ra_z$. 
Applying the Cauchy-Schwarz inequality to 
$\La F^2 ( \mu z ) \Ra_z/\La F''(\mu z)\Ra_z$ then yields that the optimal
choice is $F(x) = x^2 - \mu^2$.
For this choice the eigenvalue
$\lambda_0$ of the correlation matrix $C^P$ vanishes for large $P$.
So the optimal $F$ maximizes the signal to noise ratio between the 
eigenspace of difference vectors we want to estimate
and the orthogonal space.

For the activation function $h(x) = \erf(x)$ one finds that
$\alcK = 1.96 K$ when $F$ is optimal, whereas the simple
choice $F(x)=x^2$ yields a critical $\alpha$ which
is one and a half times higher. 
Simulation results for $K=7$ plotted in Fig. 2 show that
the large $K$ theory provides a reasonable approximation already for
networks with quite few hidden units.

To round off the analysis of the regression problem, we obtain a theoretical
prediction for the generalization error achieved by combining the two stages 
of our procedure. A simple calculation shows that the overlap
$r$ of the Hebbian vector $\bar{J}^\mu$ with $\bar{B}$, 
$r =\bar{B}^T\bar{J}^\mu /|\bar{J}^\mu|$, for large $N$ satisfies 
$r = (1 + \frac{\arcsin(2/3)}{2 \alpha K})^{-1/2}$.
Further, using the explicit expression for $\epsg$ given
in \cite{Saa95} and the values $r$ and $\rho$, we can
calculate the minimum of $\epsg$ in the restricted space and
find a theoretical prediction for the generalization
error obtained by the second stage.
This prediction is
compared to the simulations in Fig. 1 and 2.

We next consider classification problems, that is we assume that the output
function of the network given by Eq. (\ref{comm}) is $g(x) = \sign(x)$. 
Then, since the output is binary,
$\lamorth = \lamav$ holds for any choice
of the weight function $F$, and our procedure cannot immediately be 
applied. However, it
is possible to gain information about the target rule by not just 
considering its output $\tau^\mu$ but by comparing the output to the value
$\bar{B}^T\xi^\mu$ which would have been obtained by the linearized network,
$g(x)=h(x)=x$. This is feasible since, even for classification, $\bar{B}$ can
be estimated by the Hebbian vector $\bar{J}^P$ defined by (\ref{hebb}).
We are thus lead to consider more general correlation matrices of
the form
$
C^P = P^{-1}\sum_{\mu = 1}^P 
      F\left(\tau(\xi^\mu),\mbox{$\bar{J}^{\mu-1}$}^T\xi^\mu\right) 
      \xi^\mu {\xi^\mu}^T\,.
$
A reasonable way of choosing $F$ is to focus on
inputs $\xi^\mu$ where the target output has a different sign than
the output of the linearized network. So we consider
$F(x,y) = \Theta(-xy)-\mu$, where $\Theta$ is the Heavyside step function.

In the large $P$ limit, the matrix $C^P$ has the same eigenspaces as in
the case of regression and the three eigenvalues will in general be different.
For the activation function we chose $h(x) = \sign(x)$ to compare our results 
with the findings for  on-line Gibbs learning \cite{Kim98}, to our knowledge 
the only algorithm which has been simulated in any detail for classifications
task with a connected committee machine.
For this $h$ one finds 
$\lamdiff > \lamorth > \lamav$, and for large $K$ to leading order
$\lamdiff -  \lamorth = \sqrt{2\pi-4}/(\pi^2 K)$. Numerical simulations of 
the procedure are shown in Fig. 3 using $\mu = \pi^{-1} \arccos\sqrt{2/\pi}$.
Motivated by our findings in the case of regression,
this choice of $\mu$  yields $\lamorth = 0$ for large $K$. 
While the training sets needed to achieve
good generalization are much larger than for regression, they are, 
for exactly the same architecture, approximately $20$-times smaller than
for on-line Gibbs learning\cite{Kim98} already for $N=150$.

Throughout this Letter we have assumed that the inputs are drawn
from an isotropic distribution.  Typically, in practical applications, 
the input data itself will have some structure,
and in this case the inputs have to be whitened before applying our procedure.
To this end one computes the correlations of the inputs by
$D^P = P^{-1}\sum_{\mu=1}^P \xi^\mu {\xi^\mu}^T$. Then instead of just
considering
$C^P$, one finds the extremal eigenvectors $\tilde{\Delta}_j$
of the transformed matrix ${R^{-1}}^TC^PR^{-1}$, where the square matrix
$R$ is the Cholesky factorization of $D^P$, $R^TR = D^P$. Then an estimate
for the space spanned by the difference vectors is obtained by transforming
back, setting $\Delta_j = R^{-1} \tilde{\Delta}_j$. 
Numerical simulations for $F(x)=x^2$ show that whitening noticeably improves
the performance of our algorithm even when the inputs are picked from an 
isotropic distribution. This is due to the fact that
the spectrum of the empirical correlation matrix $D^P$ has a finite width
even when it converges to a delta peak with increasing $P$. 

In summary, we have presented the first learning algorithm for a realistic
multilayer network which can be exactly analyzed in the thermodynamic limit 
and yields
good generalization already for a finite ratio of training examples to free
parameters. This contrasts with the behavior of traditional
on-line procedures \cite{Saa95,Bie96} for target functions such as 
the ones considered in this Letter. As long as there are
on the order of $N$ on-line steps,  the dynamics is exactly described
by deterministic differential equations for a  set of self-averaging order 
parameters \cite{Ree98} in the limit of large $N$. 
However, even when tuned by choosing a good learning rate,
the dynamics is stuck in a badly generalizing plateau on this 
time scale unless some information about the target function is built into
the initial conditions. For the realistic case of randomly chosen
initial values,  the symmetries are eventually broken 
due to  small initial fluctuations and the 
accumulation of fluctuations during the dynamics, but
only when the number of on-line steps is much larger than $N$.
This symmetry
breaking is not adequately described by the deterministic differential 
equations. When the training set is sampled without replacement,
this divergence of the time scale  means that large 
numbers of examples are needed. But even in the recently analyzed
scenario of sampling with replacement \cite{Coo00b}, 
the above theoretical problems remain and are compounded since
the analysis requires involved techniques such as dynamical 
replica theory and quite a few approximations 
already on the time scale $N$.

Hence, we believe that the findings in this Letter 
open new avenues for research and
that in particular the performance of the algorithm for classification tasks
deserves a more detailed analysis. But the perhaps most intriguing  aspect of
our procedure is that it does not really assume that the number 
of hidden units in the architecture is known a priori. This is 
highlighted by the inset of Fig. 2 showing that the  number 
of hidden units can be determined by inspecting the eigenvalue spectrum of 
the correlation matrix. 
So our findings also provide a novel perspective on the problem
of model selection.

The work of two of us (C.B. and R.U.) was supported by the 
Deutsche Forschungsgemeinschaft. 

\nocite{Urb96}

\begin{figure}[p]
    $\begin{tabular}{ccc} \large$\rho$&
        \mbox{\begin{tabular}{c}\includegraphics[scale=0.4,clip]{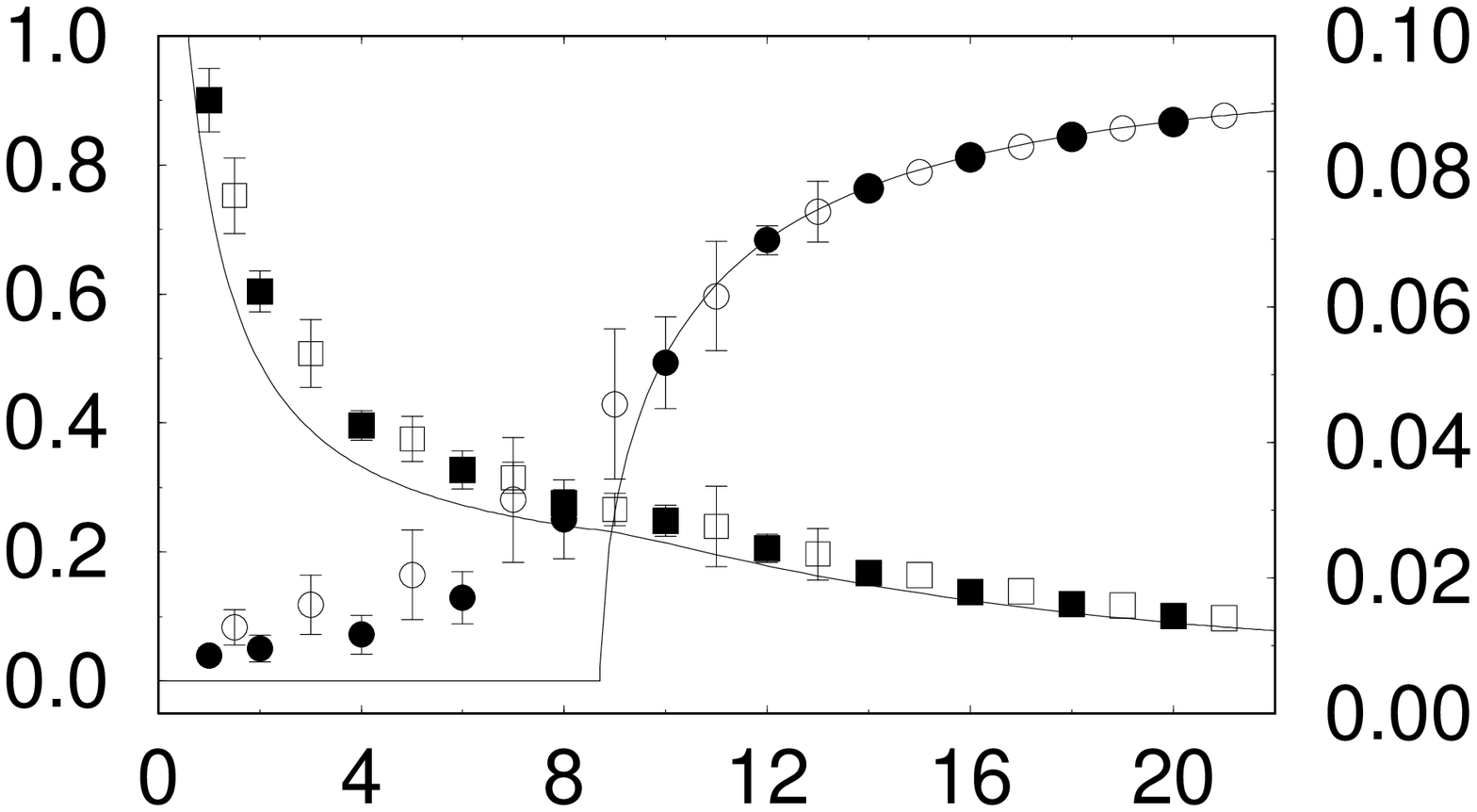}
              \end{tabular}}
        & \large$\epsg$\\
        &\large$\alpha$
   \end{tabular}$   
 \caption{ Results for $K=3$. The evolution of $\rho$ for 
 $N=400$ ({\large$\circ$}) and $N=1600$ ({\large$\bullet$}). 
 The right axis relates to the 
 values of $\epsilon_g$ found when the two stages of our procedure are
 combined,  $N=400$ ($\scriptstyle\square$) and $N=1600$ 
 ($\scriptstyle\blacksquare$).  For
 $\epsilon_g$ the value of $\alpha$ refers to the number of examples
 in both training sets, $\alpha = (P+\hat{P})/KN$. The full lines show the 
 theoretical prediction for the thermodynamic limit. Where
 not shown, errorbars are smaller than the symbol size. 
 }
\end{figure}

\begin{figure}[p]
   \begin{tabular}{ccc} \large$\rho$&
        \mbox{\begin{tabular}{c}\includegraphics[scale=0.4,clip]{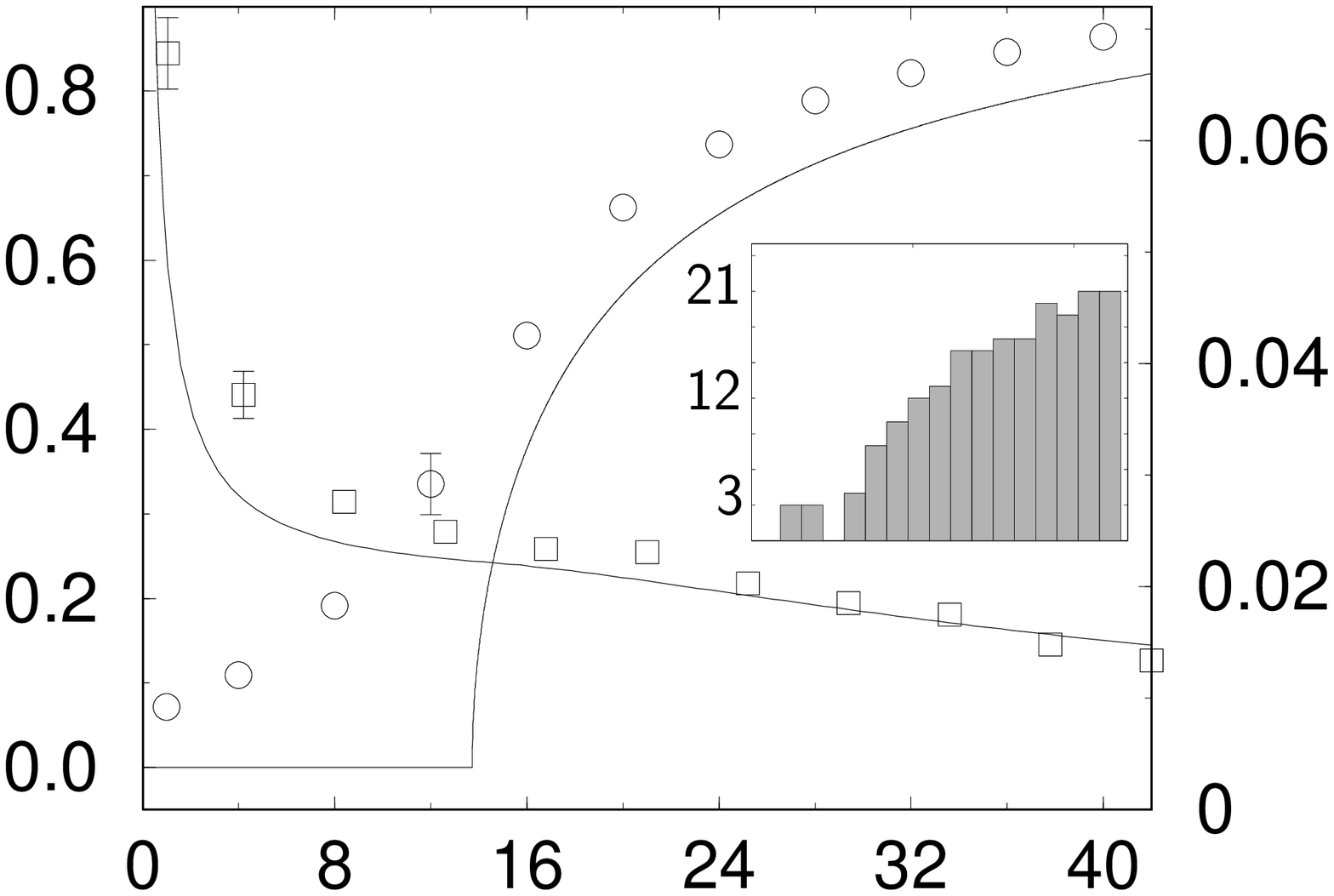}
              \end{tabular}}
        & \large$\epsg$\\
        &\large$\alpha$
   \end{tabular}   
 \caption{ Results for $K=7$ using the weight function
   $F(x)= x^2-\mu^2$.  The numerical simulations for $N=2000$
   are compared to the theoretical curves found in the large
   $K$ limit. Where not shown, errorbars are smaller than the
   symbol size. The inset shows a histogram of the 200 smallest
   eigenvalues of $C^P$ for a single training set at $\alpha =
   22$. A
   gap separates the 6 smallest eigenvalues from the rest of
   the spectrum. The range of the shown eigenvalues is
   $[-0.1,-0.07]$.
 }
\end{figure}

\begin{figure}[p]
   \begin{tabular}{ccc} \large$\rho$&
        \mbox{\begin{tabular}{c}
           \includegraphics[scale=0.375,clip]{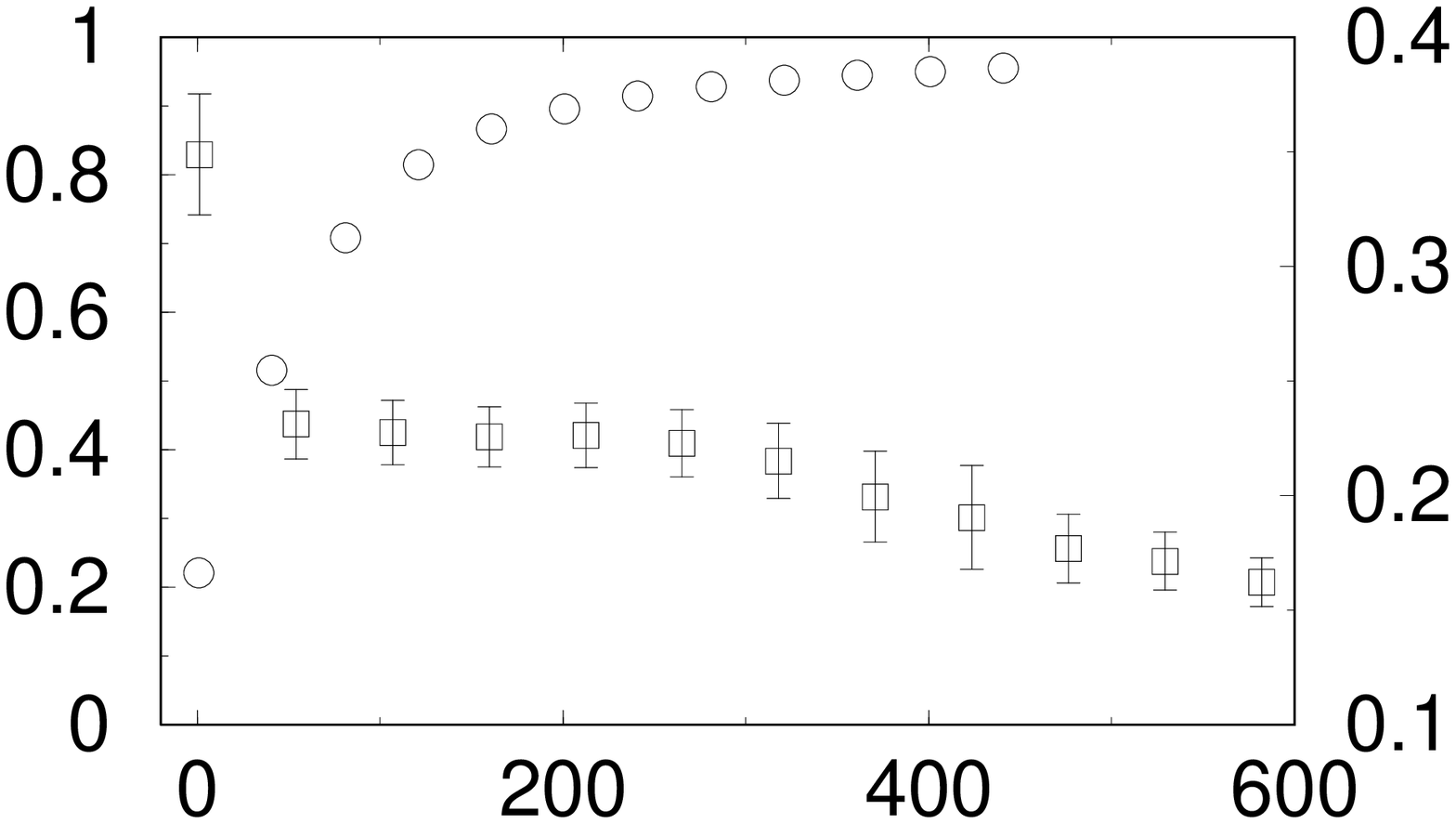}
              \end{tabular}}
        & \large$\epsg$\\
        &\large$\alpha$
   \end{tabular}   
 \caption{Numerical simulations of the classification
   problem for $K=7, N=150$ showing the values of $\rho$
   ({\large$\circ$}) and $\epsg$ ($\scriptstyle\square$). The
   errorbars for the $\rho$ values are smaller than the symbol
   size. Here $\epsg$ is the probability of misclassifying a
   new input. Training in the restricted space during the
   second stage of our procedure uses the variant of the
   perceptron learning rule described in \protect\cite{Urb96}
   for tree committee machines.
   }
\end{figure}

\end{document}